\documentstyle[floats,twocolumn,aps,prl]{revtex}
\begin{document}
\ifpreprintsty\else
\twocolumn[\hsize\textwidth%
\columnwidth\hsize\csname@twocolumnfalse\endcsname
\fi
\title{Theory of the Half-Polarized Quantum Hall States}
\author{V.M. Apalkov$^\ast$, T. Chakraborty$^\ast$, 
P. Pietil\"ainen$^\dagger$, and K. Niemel\"a$^\dagger$}
\address{$^\ast$Max-Planck-Institut f\"ur Physik Komplexer Systeme,
Dresden, Germany \\
$^\dagger$Theoretical Physics, University of Oulu, Finland}
\maketitle
\begin{abstract}
We report a theoretical analysis of the half-polarized quantum
Hall states observed in a recent experiment. Our numerical results
indicate that the ground state energy of the quantum Hall $\nu=\frac23$ 
and $\nu=\frac25$ states versus spin polarization has a downward 
cusp at half the maximal spin polarization. We map the two-component 
fermion system onto a system of excitons and describe the ground state 
as a liquid state of excitons with non-zero values of exciton angular 
momentum.
\end{abstract}
\ifpreprintsty\clearpage\else\vskip1pc]\fi
%\pacs{73.40.Hm,71.45.Gm,73.20,Dx}
\narrowtext

In recent years it has become increasingly clear that spin 
degree of freedom plays an important role in the fractional quantum 
Hall effect (FQHE) where many novel and interesting spin-related 
phenomenon have been observed both theoretically and experimentally 
\cite{book}. One of the important problems in this field is the 
influence of Zeeman splitting on the properties of FQHE systems, 
in particular, on the ground state spin-polarizations. It is now well 
established that for some filling factors ($\nu=\frac1m$) the ground 
state is fully spin-polarized for all values of Zeeman splitting, 
while for other filling factors (for example, $\nu =2/3$, $2/5$) the
ground state is fully polarized only for large values of Zeeman
splitting but unpolarized or partially polarized for small (or zero)
values of Zeeman energy\cite{book,rev}. One interesting problem then
is to find the state for intermediate values of Zeeman energy. That 
problem was highlighted in a recent experimental work \cite{kukushkin}, 
where the magnetic field driven spin transitions at various FQHE 
states were reported and in particular, at $\nu=\frac23$ and
$\nu=\frac25$, weak features (plateau-like singularities) were observed 
at half the maximal spin polarization of the system. Observed stability 
of the half-polarized states means that the ground state energy of the 
system as a function of spin polarization should have non-monotonic 
behavior at half polarization. Our earlier work \cite{karri} did not 
provide much information about the nature of states at 
half-polarization. In this paper, we have explored possible 
ground states at half-polarization for filling factors $\nu=\frac25$
and $\nu=\frac23$. We find that the ground state energy versus spin
polarization has a downward cusp at half polarization that might 
describe stability of the observed state.

The FQHE system at filling factor $\nu=\frac25$ can be 
described as a composite fermion (CF) system with total filling 
factor $\nu=2$ \cite{cfermion}. At high values of the Zeeman energy 
the composite fermions will occupy $n=0$ $\uparrow$-spin and 
$n=1$ $\uparrow$-spin Landau levels of composite fermions. 
As a result we have a fully spin-polarized  state. 
At low Zeeman energies they will occupy $n=0$ 
$\uparrow$-spin and $n=0$ $\downarrow$-spin Landau levels
which will result in an unpolarized state. At intermediate 
values of Zeeman energy the composite fermions will 
fully occupy $n=0$ $\uparrow$-spin Landau level and partially
occupy $n=0$ $\downarrow$-spin and $n=1$ $\uparrow$-spin 
levels with filling factors $\nu_1$ and $\nu_2$ respectively,
with $\nu_1+\nu_2=1$. The half polarized state corresponds to 
$\nu_1=\nu_2=\frac12$. As the fully occupied state can 
be considered as a non-dynamical background, the composite 
fermions occupying the partially filled levels can be thought of 
as a system consisting of two types of fermions with $\nu=1$. 
The Hamiltonian of the two-component system has the form
\begin{eqnarray}
\nonumber
{\cal H}= \frac12 \sum_{\alpha \beta} 
            && \int \int d\vec{r}_1 d\vec{r}_2 
 V_{\alpha\beta}\left(\left|\vec{r}_1-\vec{r}_2\right|\right)\\
 &&\times \psi^{+}_{\alpha} (\vec{r}_1) \psi^{+}_{\beta} (\vec{r}_2) 
          \psi_{\beta} (\vec{r}_2) \psi_{\alpha}(\vec{r}_1)
%  \nonumber
\end{eqnarray}
where $\alpha, \beta=1$ and 2, $V_{11}$ is the interaction potential
between fermions of type 1,  $V_{22}$ is the interaction potential
between fermions of type 2,  and $V_{12}$ is the potential 
between fermions of types 1 and 2.
The specific feature of this system is that the Hamiltonian is 
completly ``non-symmetric'', i.e., all interaction potentials $V_{11}$, 
$V_{22}$ and $V_{12}$ are different.   
 
Alternatively, we can also describe the $\nu=\frac25$ state as the 
daughter state of the $\nu=\frac13$ system \cite{book}. Then the 
spin-polarized state of the $\nu=\frac25$ system is due to 
condensation of spin-polarized quasiparticles with filling factor 
$\nu=\frac12$ and the unpolarized state as the condensation of 
spin-reversed quasiparticles \cite{kane}. For intermediate polarization 
we have the system of spin-polarized and spin-reversed quasiparticles 
with $\nu=\frac12$. Because they are Bose particles we can 
map this system into the system of fermions with $\nu=1$. Here again, 
as for the composite fermion picture, we have two types 
of fermions with Hamiltonian (1). We also have the similar picture 
for $\nu=\frac23$ state, which can be described as the daughter state 
of $\nu=1$ system with condensation of spin-polarized and spin-reversed 
holes of $\nu=1$.  

For strong enough repulsion between the fermions at the same point 
we expect that the ground state of the 
system at $\nu_1=\nu_2 =\frac12$ and other values of $\nu_1$
to be the Halperin-(1,1,1) state \cite{book,bert}
\begin{eqnarray*}
\psi=&&\prod_{i<j=1}^{N_1}(z_i-z_j)\prod_{k<m=1}^{N_2}
(\tilde z_k-\tilde z_m)\prod_{i=1}^{N_1}\prod_{k=1}^{N_2}
(z_i-\tilde z_k)\\
&&\times\prod_{i=1}^{N_1}{\rm e}^{-\vert z_i\vert^2/4\ell_0^2}
\prod_{k=1}^{N_2}{\rm e}^{-\vert\tilde z_k\vert^2/4\ell_0^2}
\end{eqnarray*}
where $N_1$ is the number of fermions of type 1, $N_2$ is the 
number of fermions of type 2;  $N_1+N_2=N$. Here $z_{j}$ are the 
coordinates (complex) of fermions of type 1, $\tilde z_{j}$ are 
those of fermions of type 2. If we consider the system of two 
types of fermions as a two-level system and introduce a pseudo-spin 
$\tau$ for the states at different levels then the Halperin-(1,1,1) 
liquid state has $\tau=N/2$ and $N/2 \ge \tau_z \ge -N/2$. If this 
state is the correct ground state then the transition from a polarized 
state to an unpolarized state of the system is just the rotation
of pseudo-spin vector from $\tau_z =-N/2$ to $\tau_z =N/2$ with
fixed value of the total pseudo-spin $\tau=N/2$. But  then
the ground state energy of the system is monotonic with
polarization of the system (quadratic function) without
any singularity at half polarization. This means that the Halperin
state is not the true ground state of the half-polarized state.

\begin{figure}
\begin{center}
\begin{picture}(120,130)
\put(0,0){\includegraphics{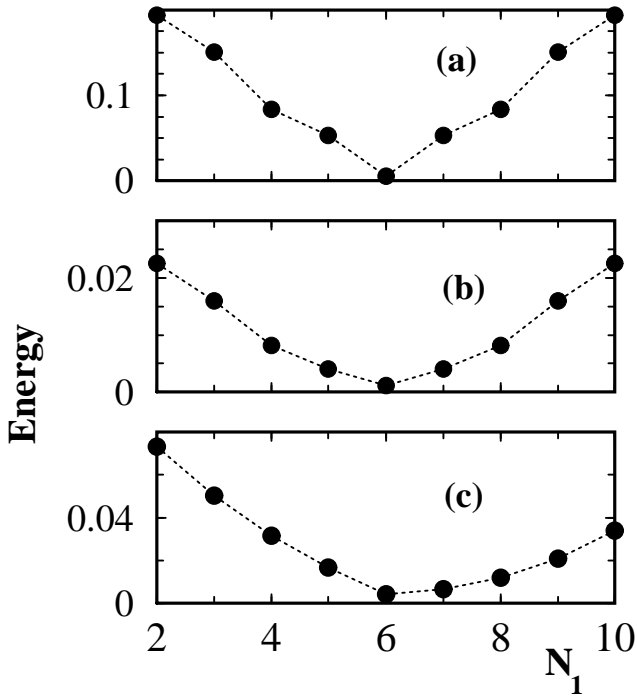}}
\end{picture}
\vspace*{4.8cm}
\caption{Ground state energy as a function of filling factor of 
fermions of type 1, $\nu_1=N_1/12$, for (a) ``symmetric'', 
(b) ``quasiparticle'' and (c) ``quasihole'' systems. The energy is in 
units of $e^2/\epsilon\ell_0$.
}
  \end{center}
\end{figure}

In Ref.\cite{murthy} it was proposed that the half-polarized ground 
state is a charge-density-wave (CDW) of composite fermions. To check 
this claim we have compared the energies of proposed CDW state and 
the Halperin-(1,1,1) state \cite{vadim}. The CDW of Ref.\cite{murthy} 
in our notations is formed by type-1 and type-2 fermions on a square 
lattice. We calculate the cohesive energy of this state from 
\cite{fukuyama}
\begin{eqnarray}
\nonumber
&&E_{\rm CDW}=-\frac12 \sum_{\vec{Q}} \left(  
    V^{\rm HF}_{11}(\vec{Q})\Delta_{1}(\vec{Q}) \Delta_{1}(-\vec{Q})
    \right. \\
    \nonumber
&&+ V^{\rm HF}_{22}(\vec{Q})\Delta_{2}(\vec{Q})\Delta_{2}(-\vec{Q})
+ \left. 2 V^{\rm H}_{12}(\vec{Q})\Delta_{1}(\vec{Q})\Delta_{2}(-\vec{Q})
  \right)\\
\end{eqnarray}
where $V^{\rm HF}_{11}(\vec{Q})$ and $V^{\rm HF}_{22}(\vec{Q})$ are 
the Hartree-Fock potentials for fermions of type 1 and 2, respectively, 
$V^{\rm H}_{12}(\vec{Q})$ is the Hartree interaction potential between 
fermions of types 1 and 2. The order parameter $\Delta_{\alpha}(\vec{Q})$ 
of the CDW corresponding to wave vector $\vec{Q}$ for fermions of 
type $\alpha $ is taken to be non-zero only for reciprocal vectors: 
$\vec{Q}=(\pm Q_0,0),(0,\pm Q_0)$ and $(\pm Q_0,\pm Q_0)$, 
$(\pm Q_0,\mp Q_0)$, where $Q^2_0\ell_0^2=\pi$ ($\ell_0$ is the 
magnetic length for composite fermions). The energy of the 
Halperin-(1,1,1) state is calculated from
\[
E_{(1,1,1)}=-\frac1{4\pi}\int d^2 r V_{\rm eff}(r)[g(r)-1],
\]
where $g(r)=1-\exp(-r^2/2\ell_0^2)$ is the correlation function 
of a fully occupied Landau level and 
\[
V_{\rm eff}(r)=\frac14\left[V_{11}(r)+V_{22}(r)+2V_{12}(r)\right]
\]
is the effective interaction between composite fermions.
 
We have considered Coulomb interaction between composite fermions
of types 1 and 2. The interaction asymmetry here results from their 
different form factors because fermions belong to different Landau 
levels. The energy of the Halperin liquid state at half polarization 
is $-0.196 e^2/\varepsilon \ell_0$ and is {\it lower} than that of 
the proposed CDW state, $-0.123 e^2/\varepsilon \ell_0$.
This means that the proposed CDW state is not the lowest energy state
and therefore not the ground state at half-polarization.

To find the true ground  state of the system that has lower energy 
than the Halperin state and that can also explain the half-polarized 
singularity, we considered the system described by the Hamiltonian 
(1) numerically in a spherical geometry \cite{haldane,fano}. All 
computations were done for a 12 fermion system on a sphere with sphere 
parameter $q=5.5$, where the radius of the sphere 
$R=\sqrt{q}=2.34$ in units of magnetic length $\ell_0$. Because the 
Hamiltonian (1) does not change the number of fermions of a given 
type the eigenstates of the system can be classified by the number 
of fermions of type 1 ($12\ge N_1\ge 0$) and by angular momentum $L$ 
(due to spherical geometry). In what follows, we have investigated 
three systems: 
(i) a ``symmetric'' system $(V_{11}=V_{22}=V_{12})$, where the
interaction potentials are Coulombic except at the origin where it
is taken to be less repulsive, 
(ii) the nonsymmetric system where the interactions between fermions 
were taken as the interactions between quasiparticles of $\nu=\frac13$ 
state -- the ``quasiparticle'' system. These interaction potentials 
were found from finite size computations following method of 
Ref. \cite{morf}, and, (iii) the nonsymmetric system where the 
interactions between fermions were taken as the interactions between 
quasiholes of $\nu =1$ state -- the ``quasihole'' system.  
For all these systems we found a similar behavior: The ground 
state energy as a function of filling factor $\nu_1$ ($N_1$) has 
a {\it downward cusp} at $\nu_1=\frac12$ ($N_1=6$), as seen in
Fig. 1. This point corresponds to a half-polarized state of the 
original system and the cusp indicates stability of the observed 
\cite{kukushkin} half-polarized state.

It should be mentioned that for ``quasiparticle'' and ``quasihole'' 
systems we do not include in the Hamiltonian (1) the different 
creation energies for polarized and spin-reversed quasiparticles 
and quasiholes, which acts as an effective internal Zeeman splitting 
and makes the unpolarized state ($\nu_1=1$) the ground state at 
zero value of real Zeeman energy. These terms are monotonic functions 
of polarization and do not change the singular behaviour at half 
polarization.

In order to analyze the symmetric system more carefully,
we consider our two-component fermion system with $\nu=1$ 
as a system of excitons, i.e., the electron-hole pairs. We choose 
type-1 fermions as electrons and the absence of type-2 fermions as holes. 
In this case the filling factors for electrons and holes are the same and 
is equal to $\nu_1$, which is also the exciton filling factor. 

\begin{figure}
\begin{center}
\begin{picture}(100,130)
\put(0,0){\includegraphics{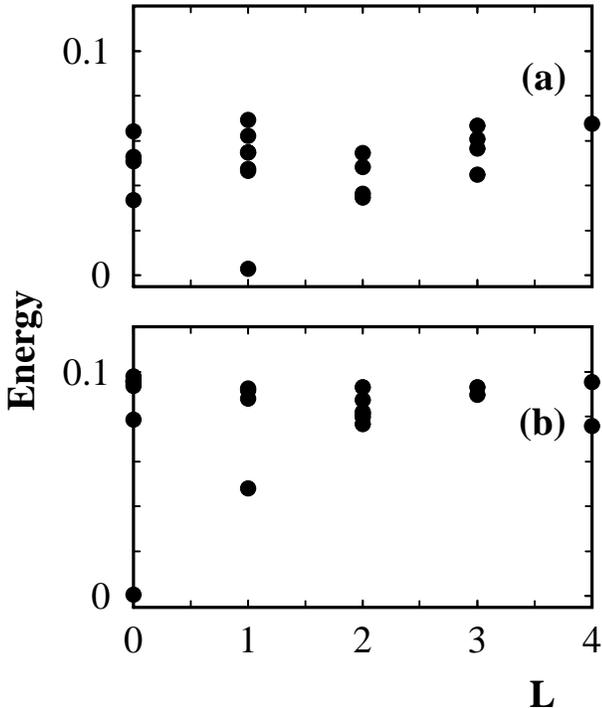}}
\end{picture}
\vspace*{5.3cm}
\caption{Energy spectrum of the symmetric system for (a) $N_1=5$ 
and (b) $N_1=6$. Energy is in the units of $e^2/\epsilon\ell_0$. 
}
  \end{center}
\end{figure}

It is well known that for a symmetric (Coulomb) multi-exciton system 
the ground state is a Bose-condensed state of excitons with zero 
momentum \cite{lerner,rashba,lozovik,rice}. In original fermion 
language these
states are the Halperin states. For Coulomb interaction, these 
Bose-condensed states are the ground states
for all values of filling factor $\nu_1$ ($1>\nu_1>0$). Let us now 
decrease the repulsion between original fermions at the origin. 
We do it numerically by decreasing the value of the pseudopotential 
with zero angular momentum, $V_0$. A decrease of $V_0$ by about 40\% 
results in ground states of the multi-exciton system that are not 
the Bose-condensed states of excitons with $L=0$ (Fig. 1). These transitions 
are also accompanied by transition of the ground state of the one-exciton 
system to the state with non-zero angular momentum, $L=1$. 

Interestingly, the transition from a Bose-condensed state of 
zero-momentum excitons to a new state was also observed in a double 
layer system \cite{tapash} with $\nu=1$ when the 
separation between the layers is increased beyond some critical value 
\cite{fertig,MacDonald}. At a critical layer separation of this system, 
the dispersion relation of the collective mode becomes negative at 
momentum $q\sim 1.3/\ell_0$ and it was proposed that this transition 
is the transition to a charge density wave state \cite{quinn}. 
In our case the dispersion relation of the collective mode of the
Bose-condensed state also becomes negative but for much smaller momentum
$q\sim L/R\sim 0.4/\ell_0$, which results from the fact that the ground 
state of one-exciton system has angular momentum $L=1$. 

As stated above,
for the Coulomb interaction the ground state of the one-exciton system has  
zero momentum and the multi-exciton ground state is the Bose-condensed 
liquid state of excitons with zero momentum \cite{rice}. In analogy 
with the Coulomb system, we describe our multi-exciton ground state 
as a liquid state of excitons with non-zero angular momentum $L=1$,
a {\it non-symmetric exciton liquid}. 
Our multi-exciton system does not contain $L=0$ excitons because they 
are non-interacting and removing one such exciton would not change the 
energy of the multi-exciton system. However, in Fig. 1, the energy of 
our 5-exciton system is higher then the energy of the 6-exciton system.
That means the 6-exciton system can have only excitons with non-zero 
angular momemtum. 

In Fig. 2, the energy spectrum of multi-exciton symmetric system is 
shown for 5- and 6-exciton systems. We can see that the ground state 
of the 6-exciton system has zero angular momentum, while the ground state of
the 5-exciton system has angular momentum $L=1$. For the Halperin liquid each 
exciton effectively occupies only one state (electron and hole are in 
the same place) and the filling factor of the exciton is equal to the 
filling factor of electrons, $\nu_1$. In the non-symmetric exciton
liquid, each exciton occupies effectively two states. As a result, the 
effective filling factor of  excitons is $2\nu_1$. Then for 
$\nu_1=\frac12$ the filling factor of excitons is 1 which means that 
we have completely occupied Landau level. Results of Fig. 2 do support 
this contention because removing one exciton with angular momentum 
$L=1$ from 6-exciton system left the exciton hole with the same angular 
momentum, which means that the system behaves as though the levels 
are completely filled. The energy spectrum of the 6-exciton system, 
which corresponds to a half-polarized state of the original system, 
also has a gap. However, from the finite size results we can not 
ay with certainty if this gap will survive in the thermodynamic limit.

\begin{figure}
\begin{center}
\begin{picture}(100,130)
\put(0,0){\includegraphics{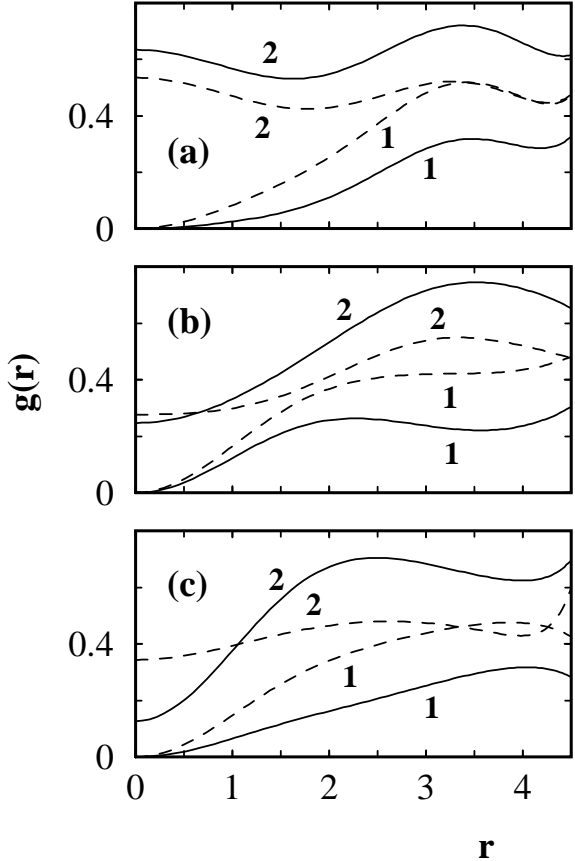}}
\end{picture}
\vspace*{7.0cm}
\caption{Ground state pair correlation functions $g_{11}(r)$
(curves labelled ``1'') and $g_{12}(r)$ (curves labelled ``2'')
for (a) symmetric, (b) ``quasiparticle'' and (c) ``quasihole'' 
systems. The solid lines are for $N_1=4$ systems, and dashed lines 
are for $N_1=6$ systems. Correlation functions are shown in 
units of maximum electron density, and $r$ is in units of the 
magnetic length.
}
  \end{center}
\end{figure}

In Fig. 3, the pair correlation functions $g_{11}(r)$ and $g_{12}(r)$ 
are shown for $N_1=4$ and 6, for symmetric (a), ``quasiparticle'' (b) 
and ``quasihole'' (c) systems. The general feature of all these 
correlation functions is the non-zero value of $g_{12}(0)$.
For the Bose-condensed state of excitons with zero momentum 
(the Halperin state) we would expect $g_{12}(0)=0$, i.e., 
the holes in the multi-exciton picture are sitting exactly at the 
position of the electrons. The non-zero values of $g_{12}(0)$ can 
therefore be directly associated with the cusp at $N_1=6$.

In closing, we have investigated the possible ground states at half
the maximal spin polarization for $\nu=\frac25, \frac23$ FQHE
states. Our results indicate that for the systems studied here,
there is a downward cusp at half polarization that reflects the
observed structure in a recent experiment \cite{kukushkin}. We 
interpret this result as due to condensation of a non-symmeric 
exciton liquid.

Two of us (V.M.A. and T.C.) would like to thank Peter Fulde for 
his support and kind hospitality in Dresden.

\end{document}